\documentclass[a4paper]{article}

\usepackage[T1]{fontenc}
\usepackage[utf8x]{inputenc}
\usepackage[english]{babel}

\usepackage[colorlinks=true, allcolors=blue]{hyperref}

\newcommand{\email}[1]{\href{mailto:#1}{\tt{\nolinkurl{#1}}}}
\newcommand{\orcid}[1]{ORCID: \href{https://orcid.org/#1}{\tt{\nolinkurl{#1}}}}

\usepackage[sfdefault,lf]{carlito}
\usepackage[parfill]{parskip}

\usepackage{fancyhdr}
\usepackage{natbib}
\usepackage{authblk}
\usepackage[a4paper,top=3cm,bottom=2cm,left=3cm,right=3cm,marginparwidth=1.75cm]{geometry}

\usepackage{amsmath}
\usepackage{graphicx}
\usepackage{booktabs}

\usepackage[colorinlistoftodos]{todonotes}

\title{The Emergence of Astroparticle Physics: From Cosmic-Ray Physics to a new Scientific Field\footnote{This article substantially expands both the historical scope and the historiographical interpretation of the paper presented at the 4th International Symposium on the History of Particle Physics which was held at CERN in Geneva from 10 to 13 November 2025. It will be published in the corresponding proceedings as follows: L. Bonolis, The Transformation of Cosmic-Ray Physics and the Emergence of Astroparticle Physics, 2027.}}
\author[1]{Luisa Bonolis}
\affil[1]{Max Planck Institute for the History of Science, Berlin; \orcid{0000-0003-3333-2135}}
\affil[*]{Corresponding author: \email{luisabonolis1@gmail.com}}
\date{}

\begin{document}
\maketitle
\thispagestyle{fancy}

\begin{abstract}
Astroparticle physics emerged during the late twentieth century as a new interdisciplinary field at the intersection of particle physics, astrophysics, and cosmology.  This article examines the historical mechanisms through which it took shape as a distinct scientific field, focusing on the progressive interaction of previously separate experimental cultures, research traditions, and scientific communities. It argues that astroparticle physics emerged not through the simple convergence of established disciplines, but through the gradual reorganization of high-energy research around scientific problems that transcended traditional disciplinary boundaries. Three broad historical trajectories shaped this process: the transformation of postwar cosmic-ray physics; the emergence of a high-energy Universe through relativistic astrophysics and new astronomical windows; and the growing interaction between particle physics and cosmology. Their progressive interaction transformed naturally occurring particles into astrophysical messengers and the Universe itself into a laboratory for investigating fundamental physics beyond the reach of terrestrial accelerators. At the same time, new experimental practices, international collaborations, scientific forums, and institutional structures progressively gave organizational form to this emerging research landscape. More broadly, the emergence of astroparticle physics illustrates how new interdisciplinary fields develop through the co-evolution of scientific questions, experimental cultures, research communities, and institutions.
\end{abstract}

\section*{Introduction}
\label{Intro}

Astroparticle physics emerged during the late twentieth century as a new interdisciplinary field at the intersection of particle physics, astrophysics, and cosmology. By combining observations of cosmic rays, neutrinos, electromagnetic radiation across a broad range of wavelengths, gravitational waves, and other cosmic messengers, it has opened new ways of investigating both the Universe and the fundamental laws of physics. Its historical emergence, however, cannot be understood simply as the convergence of previously independent disciplines. Rather, it resulted from a long process during which previously distinct experimental traditions, research communities, scientific questions, and institutional structures progressively became interconnected.

Existing historical and philosophical studies have provided important perspectives on the formation of astroparticle physics, through historical reconstructions, accounts by field's protagonists, and analyses of its epistemological organization and pragmatic coherence \citep{Cirkel-Bartelt:2008fk} \citep{Falkenburg:2012aa} \citep{Falkenburg:2012ab} \citep{Falkenburg:2014} \citep{Bettini2025}. The broader transformation of twentieth-century astrophysics and cosmology has been comprehensively reconstructed by Malcolm Longair in \textit{The Cosmic Century}, one of the most comprehensive historical synthesis of the transformations that reshaped modern astronomy, high-energy astrophysics, and cosmology during the twentieth century \citep{Longair2006}. More specifically, the emergence of astroparticle physics as a distinct field has been examined from both historical and philosophical perspectives by Cirkel-Bartelt and by Rhode and Falkenburg, who have emphasized its interdisciplinary character and the gradual reconfiguration of disciplinary boundaries. Building upon these contributions, this article adopts a different historiographical perspective. Rather than focusing primarily on the disciplinary identity of astroparticle physics, it reconstructs the historical mechanisms through which previously distinct experimental cultures, research traditions, and scientific communities progressively became interconnected through common scientific challenges, complementary observational strategies, and increasingly shared research infrastructures.

Within this broader historical process, three trajectories proved particularly significant. 

\begin{itemize}
\item The first was the transformation of postwar cosmic-ray physics. As accelerator-based particle physics assumed the leading role in the discovery of elementary particles, cosmic-ray research progressively shifted its emphasis toward the origin, acceleration, and propagation of high-energy particles, developing distinctive experimental practices and geographically distributed infrastructures that would later provide important foundations for astroparticle physics. 

\item The second trajectory was the emergence of a new vision of the high-energy Universe through radio, X-ray, $\gamma$-ray, and neutrino astronomy, together with the growing importance of compact objects and relativistic astrophysics. 

\item The third was the increasing interaction between particle physics and cosmology, particularly from the 1970s onward, when Grand Unified Theories, particle cosmology, and the study of the early Universe established new connections between microphysics and cosmology. 
\end{itemize}

These trajectories were not merely parallel historical developments. Each reflected a distinct transformation in the epistemic role assigned to Nature itself: naturally occurring particles became astrophysical messengers, the violent Universe became a source of extreme physical conditions inaccessible on Earth, and the early Universe became a laboratory for testing fundamental interactions. Together, these transformations progressively reshaped the scientific questions being addressed as well as the experimental practices, collaborative cultures, and institutional settings through which they were investigated.

Although analytically distinguishable, these trajectories progressively became interconnected around shared scientific problems. Their interaction fostered the epistemic, experimental, and institutional reconfiguration that ultimately gave rise to astroparticle physics as a distinct scientific field. Throughout this process, cosmic-ray physics occupied a distinctive position. It preserved an experimental culture based on naturally occurring high-energy particles, geographically distributed observatories, and the exploitation of natural environments as integral components of the experimental apparatus. As new questions concerning neutrinos, $\gamma$ rays, dark matter, proton decay, and the origin of cosmic rays emerged, previously distinct research traditions increasingly became interconnected through shared experimental challenges, complementary observational approaches, and new forms of collaboration. The resulting field owed less to the convergence of established disciplines than to the progressive organization of research around common scientific problems.

The analysis developed here suggests that the emergence of astroparticle physics can be understood through two closely related conceptual transformations.  First, naturally occurring particles were progressively reinterpreted as messengers capable of providing direct information about otherwise inaccessible astrophysical sources and environments. Second, the Universe itself increasingly came to be regarded as a natural laboratory in which physical processes beyond the reach of terrestrial accelerators could be investigated. Together, these complementary perspectives gradually reshaped the scientific questions, experimental practices, and observational strategies of high-energy research. This transformation also found expression in new scientific forums, collaborative infrastructures, and institutional initiatives. Conferences, underground laboratories, international collaborations, and specialized journals increasingly brought together communities that had previously developed along separate trajectories. By the end of the twentieth century, these interactions had acquired the coherence of a recognizable scientific field whose identity rested less on a single discipline than on a shared set of scientific questions, experimental strategies, and observational practices. 

Based on this historical reconstruction, it is argued that astroparticle physics emerged not through the simple convergence of previously established disciplines, but through the progressive reconfiguration of historically independent research traditions around shared scientific problems. This process generated new conceptualizations of Nature, common experimental cultures and research infrastructures, and ultimately a new scientific field.

The pages that follow examine this historical process from the transformation of postwar cosmic-ray physics to the institutional recognition of astroparticle physics by the end of the twentieth century. Rather than presenting a chronological survey of discoveries, they explore how previously distinct experimental cultures, theoretical traditions, and scientific communities progressively became interconnected, giving rise to a new way of investigating the high-energy Universe that laid many of the conceptual, experimental, and organizational foundations of contemporary multi-messenger astrophysics.

\section{The Transformation of Cosmic-Ray Physics and the Discovery of the High-Energy Universe}
\label{Section 1}

Long before astroparticle physics acquired an institutional identity, many of its experimental traditions, scientific questions, and research communities had already begun to take shape within postwar cosmic-ray physics. By the 1950s, cosmic rays occupied a unique position in physics. During the previous decades they had served as the principal source of information about the subatomic world, contributing to the discovery of the positron, muon, pion, and numerous other elementary particles \citep{Colloque:1982fk} \citep{Brown-Hoddeson:1983uq} \citep{Brown:1989bh}. Although the commissioning of the first high-energy proton accelerators, notably the Cosmotron (1952) and the Bevatron (1954) in the United States, rapidly shifted the discovery of elementary particles toward laboratory machines, cosmic rays retained a distinctive scientific significance \citep{Marshak:1983uq} \citep{Schweber:1989aa}. 
As James Cronin later emphasized, the 1953 conference at Bagn\`eres de Bigorre symbolized the transfer of leadership in subatomic physics from cosmic-ray studies to accelerator experiments \citep{Cronin:2011qy}. Yet cosmic rays acquired a new scientific role. They remained natural probes of energies and environments inaccessible in the laboratory, and even continued to produce particle-physics discoveries, as shown by K. Niu’s 1971 observation of charmed particles in cosmic-ray emulsions \citep{Niu:1971za}. As accelerators took over the systematic study of elementary particles, cosmic-ray research increasingly turned toward the origin, acceleration, composition, and propagation of high-energy particles in space. Because cosmic rays consist predominantly of protons and atomic nuclei of astrophysical origin, their importance lay not simply in what they were, but in the information carried by their spectra, abundances, and paths through interstellar matter. In this way, cosmic-ray physics became a bridge between particle physics and astrophysics.

The cosmic rays of extraordinary energy further reinforced their distinctive scientific status. Extensive air-shower experiments revealed that Nature is capable of accelerating particles to energies vastly exceeding those attainable in terrestrial accelerators \citep{Greisen1960}. A particularly striking example came in 1962, when John Linsley reported at Volcano Ranch an event with an energy approaching $10^{20}$ eV \citep{Linsley:1963km}. Such observations demonstrated that the most energetic phenomena in the Universe could no longer be understood simply as extensions of laboratory physics. Nature itself thus became the only available source of particles at these extreme energies, transforming the cosmos into a laboratory of fundamental high-energy inquiry \citep{Kampert:2012fk}. Understanding how and where such particles were accelerated consequently became one of the central challenges inherited by high-energy astrophysics and, later, by astroparticle physics.

Beneath many developments in cosmic-ray research lay a persistent and unresolved question: where do cosmic rays come from? Since charged particles are deflected by Galactic and intergalactic magnetic fields, their trajectories cannot be traced back to their sources. The source problem thus became one of the principal historical threads linking cosmic-ray physics with the later development of $\gamma$-ray and neutrino astronomy. While these emerging observational fields were motivated by a broad range of astrophysical questions, they shared a common objective: identifying the sites and mechanisms responsible for the acceleration of high-energy particles.

This dual scientific perspective—treating cosmic rays simultaneously as elementary particles and as probes of astrophysical environments—was reflected in a distinctive experimental culture. Unlike accelerator particle physics, which increasingly centred on dedicated laboratory facilities and artificially produced particle beams, cosmic-ray research relied on geographically dispersed experimental sites that exploited natural environments as integral components of the experimental apparatus. Mountains, balloon platforms, underground mines, deserts, polar regions, and high-altitude plateaus all became laboratories for the study of cosmic radiation. The atmosphere itself served as a giant interaction medium, transforming incoming cosmic rays into extensive cascades of secondary particles that could be observed at ground level. Physicists learned to exploit these natural processes through a flexible and technologically heterogeneous approach that combined a wide variety of detectors and observational techniques.

This way of doing physics also fostered distinctive forms of scientific collaboration. Because cosmic-ray experiments depended on access to particular geographical environments rather than on a single central machine, the field developed extensive international networks linking laboratories across continents. Scientific cooperation often emerged around shared observational opportunities rather than around common accelerator facilities. Many regions that occupied only a limited place within the geography of accelerator physics—including India, Bolivia, and later Tibet—became important centres of cosmic-ray research because of their natural advantages. In this respect, cosmic-ray physics anticipated several characteristics that would later become typical of astroparticle physics: geographically distributed infrastructures, hybrid instrumentation, international collaborations, and the exploitation of unique natural environments as scientific resources.

Far from disappearing with the rise of accelerator physics, cosmic-ray research developed into a truly international enterprise that linked diverse scientific traditions and provided many of the experimental techniques, infrastructures, and research communities that would later contribute to the emergence of astroparticle physics. This transformation found one of its clearest expressions in the scientific network built by Bruno Rossi. From his group at MIT, Rossi connected research on cosmic rays, extensive air showers, space science, and eventually X-ray astronomy, while maintaining close links with Homi Bhabha and B. V. Sreekantan in India and Minoru Oda in Japan, and many other researchers worldwide. These interactions contributed to the formation of scientific traditions that later became important components of astroparticle physics. At the same time, the Indian programme initiated by Bhabha and the Soviet research associated with Georgii Zatsepin and Aleksandr Chudakov established influential lines of investigation spanning cosmic rays, underground physics, neutrinos, and atmospheric Cherenkov techniques. Although differing in emphasis, these traditions formed part of a common international culture centred on naturally occurring high-energy particles.

One particularly significant development was the emergence of underground research. Beginning in the 1950s, deep mines and tunnels were increasingly used to investigate the penetrating component of cosmic radiation, exploiting the overlying rock to suppress the electromagnetic and hadronic components of cosmic-ray showers. Experiments at sites such as the Kolar Gold Fields in India and the deep gold mines of South Africa later led to the first detections of atmospheric neutrinos, revealing that underground laboratories could serve not only the study of cosmic rays but also the observation of new classes of particles. Significantly, physicists initially went underground not to escape cosmic radiation, but to investigate its most penetrating component. This historical inversion is revealing. From the 1970s onward, underground environments increasingly became sites for a new generation of rare-event experiments motivated by particle physics and cosmology. While these experiments often benefited from techniques and experience developed in earlier underground cosmic-ray research, they were promoted by different scientific communities and pursued distinct scientific objectives. Some evolved into permanent underground laboratories hosting successive generations of experiments, whereas others remained closely associated with specific research programmes.

While cosmic-ray physicists were investigating the origin and propagation of high-energy particles, developments in astrophysics increasingly connected cosmic phenomena with fundamental physical processes. The influential review paper by Margaret Burbidge, Geoffrey Burbidge, William Fowler and Fred Hoyle on the stellar nucleosynthesis \citep{RevModPhys.29.547} demonstrated that the origin of the chemical elements could be understood through the sequence of nuclear processes operating throughout stellar evolution and explosive stellar events, thereby establishing nuclear astrophysics as a coherent research programme. At the same time, the rapid development of radio astronomy revealed a Universe dominated by non-thermal phenomena involving relativistic particles and magnetic fields. In a series of influential studies during the late 1950s and early 1960s, Geoffrey Burbidge emphasized the astrophysical significance of relativistic particles, synchrotron radiation, and highly energetic cosmic sources, helping to establish the picture of a non-thermal high-energy Universe \citep{Burbidge1959}. Building on these developments, William Fowler and Fred Hoyle argued in early 1963 that the extraordinary energies exhibited by quasi-stellar radio sources---soon renamed quasars---could be understood as the result of intense gravitational processes and therefore required treatment within the framework of general relativity \citep{HoyleFowler1963}. The optical identification of quasars shortly thereafter confirmed the existence of these extraordinarily energetic objects and provided the immediate impetus for the first Texas Symposium on Relativistic Astrophysics held in Dallas in 1963 \citep{robinson1965quasars}. Bringing together astronomers, relativists, particle physicists, and cosmologists the meeting marked the beginning of a sustained dialogue between general relativity and high-energy astrophysics that would later become one of the intellectual foundations of astroparticle physics. 
More broadly, it illustrated a characteristic feature of the emerging high-energy Universe: newly discovered astrophysical phenomena posed scientific questions that no single discipline could answer in isolation. In this sense, it was Nature itself that compelled previously distinct scientific communities to enter into dialogue. Increasingly, it was the scientific problems posed by Nature, rather than established disciplinary traditions, that determined the communities required to address them.

The transformation of relativistic astrophysics from a largely theoretical field into an observation-driven one also provided an astrophysical framework within which Enrico Fermi's 1949 theory of cosmic-ray acceleration \citep{fermi1949origin} acquired renewed significance. As increasingly energetic cosmic objects were discovered during the 1960s, Fermi's mechanism offered a plausible explanation for how Nature might accelerate particles to the extraordinary energies observed in cosmic rays. During the 1960s and early 1970s, the discovery of quasars, the emergence of X-ray and $\gamma$-ray astronomy, and the identification of pulsars, neutron stars, and black holes profoundly transformed scientific perceptions of the cosmos. Increasingly, the Universe appeared as a realm of compact objects, relativistic plasmas, strong magnetic fields, and powerful natural accelerators. These developments also fostered the emergence of a transdisciplinary research culture that increasingly connected astronomy, astrophysics, particle physics, and cosmic-ray studies, thereby establishing many of the intellectual premises on which astroparticle physics would later develop.\footnote{For the broader emergence of this high-energy research culture during the 1950s and 1960s, see \citep{BonolisFurlan2025a}.} 

These breakthroughs were accompanied by the advent of new scientific forums that increasingly brought together previously separate research communities. Following its inauguration in 1963, the Texas Symposia on Relativistic Astrophysics rapidly became one of the principal meeting places for astronomers, relativists, particle physicists, and cosmologists interested in compact objects, high-energy phenomena, and the structure of the Universe. At the same time, the International Cosmic Ray Conferences gradually expanded their scope. Originally centred on cosmic-ray composition, air showers, and particle interactions, they increasingly included discussions of $\gamma$ rays, neutrinos, astrophysical sources, and cosmological questions. Long before astroparticle physics acquired a name, many of the communities that would later constitute the field were already encountering one another through shared scientific problems, complementary experimental approaches, and overlapping scientific interests. By the beginning of the 1970s, many of the essential ingredients of the future field were already present. Cosmic-ray physics had developed a distinctive experimental culture, geographically distributed infrastructures, and a broad repertoire of observational techniques. Relativistic astrophysics had revealed a Universe populated by compact objects and powerful natural accelerators, while underground experiments had opened new observational windows on penetrating particles and neutrinos. Yet these elements still belonged largely to partially overlapping research traditions rather than to a unified field. During the following decade, new questions arising from particle physics, cosmology, and astrophysics would progressively reshape the experimental landscape while bringing these previously distinct communities into increasingly close interaction.

\section{New Questions beyond the Accelerator}
\label{Section 2}
By the early 1970s, many of the experimental techniques that would later become central to astroparticle physics had already been pioneered. Extensive air-shower arrays, underground studies of penetrating cosmic radiation, and the first atmospheric Cherenkov experiments had demonstrated the potential of observing naturally occurring high-energy particles. During the following decade, however, these approaches were joined by new experimental programmes inspired by Grand Unified Theories, neutrino astrophysics, and the growing interaction between particle physics, astrophysics, and cosmology. Together, these complementary experimental approaches progressively shaped the experimental landscape from which astroparticle physics emerged.

One of the clearest manifestations of this transformation occurred in underground research. As discussed in the previous section, physicists had originally gone underground to investigate the penetrating component of cosmic radiation. During the 1970s, however, underground environments increasingly attracted experiments motivated by questions arising from particle physics and cosmology, particularly the search for rare phenomena such as proton decay, magnetic monopoles, and neutrino interactions. This development did not simply represent the continuation of earlier underground cosmic-ray research. It was accompanied by new scientific proposals for permanent underground research infrastructures, exemplified by projects such as Baksan \citep{Kuzminov2012} and, later, Gran Sasso \citep{Zichichi1983} \citep{Conversi1983}, as well as by the installation of increasingly ambitious experiments at existing underground sites. During the following decades, these initiatives evolved into an international network of deep underground laboratories dedicated to neutrino physics, rare-event searches, and, increasingly, astroparticle physics, profoundly transforming the experimental landscape of fundamental physics \citep{Coccia2006} \citep{Bettini2012} \citep{Bettini2014}. 

Together, these developments reflected a fundamental change in the role of the underground environment: from a convenient location for individual experiments to a scientific infrastructure designed to support long-term programmes in rare-event physics. This new experimental paradigm brought together different scientific communities around a common requirement: the exploitation of exceptionally low-background environments as integral components of the experimental apparatus, establishing deep underground laboratories as one of the defining research infrastructures of astroparticle physics \citep{Bettini2012} \citep{Bettini2023}.
 
This transformation was driven by a broader set of scientific motivations than proton decay alone. Grand Unified Theories stimulated searches for proton decay and magnetic monopoles, while advances in neutrino physics, the growing interest in solar and astrophysical neutrinos, neutrinoless double-beta decay, dark matter, and other rare or exotic phenomena increasingly pointed toward forms of fundamental physics that could not readily be explored with accelerator experiments alone. Investigating such processes required large detector masses, long observation times, and exceptionally low radioactive and cosmic-ray backgrounds. Deep underground environments therefore acquired a new scientific significance. They were no longer merely places where the penetrating component of cosmic radiation could be studied, but experimental settings in which otherwise inaccessible phenomena predicted by particle physics, astrophysics, and cosmology might be observed. This changing conception of the underground environment progressively encouraged the development of permanent research infrastructures designed to support successive generations of long-term experiments. 

Neutrinos occupied a particularly important position within this evolving landscape because they connected several previously distinct lines of research.\footnote{See contributions to the International Conference on History of the Neutrino: 1930-2018 for a general historical overviews \citep{Cribier:2019}.} Solar-neutrino experiments, initiated by Raymond Davis in the late 1960s and interpreted in close conjunction with John Bahcall's calculations of the solar neutrino flux \citep{Bahcall:2000xu}, opened a direct observational window onto the thermonuclear processes powering the Sun \citep{Kirsten2019}. At the same time, the persistent discrepancy between the predicted and observed neutrino fluxes---the solar-neutrino problem---introduced one of the most important unresolved questions in neutrino physics, prompting decades of investigations into both solar physics and the fundamental properties of neutrinos. Atmospheric neutrinos, produced by interactions of primary cosmic rays in the Earth's atmosphere, emerged from the long tradition of cosmic-ray research and were first identified in underground experiments during the mid-1960s, most notably by an international collaboration at the Kolar Gold Fields in India, involving researchers from India, Japan, and the United Kingdom, and almost simultaneously by Frederick Reines and collaborators in the deep gold mines of South Africa \citep{Kajita2010} \citep{Lipari2018}. While initially investigated within the framework of cosmic-ray physics, atmospheric neutrinos later provided a unique natural source for studying neutrino properties and ultimately played a central role in the experimental demonstration of flavour oscillations. 

At the same time, the astrophysical potential of neutrinos was beginning to receive increasing theoretical attention \citep{Spiering2012}. Building on ideas advanced by Moisei Markov at the end of the 1950s, Markov and Igor Zheleznykh proposed in 1961 the use of large volumes of natural water to detect high-energy neutrinos from astrophysical sources, introducing a fundamentally new observational strategy based on elementary particles \citep{Markov1961} \citep{Zheleznykh2006}. Parallel theoretical developments in neutrino physics---notably the emergence of the idea of neutrino mixing and oscillations, to which Bruno Pontecorvo, Georgii Zatsepin, and other Soviet physicists made fundamental contributions, expanded into quantitative frameworks by Ziro Maki, Masami Nakagawa and Schoichi Sakata---further strengthened the conceptual connections between particle physics, astrophysics, and cosmology \citep{Bilenki2016} .
The most ambitious attempt to realize Markov's vision emerged through the DUMAND (Deep Underwater Muon and Neutrino Detector) project. Initiated during the 1970s under the leadership of John Learned and collaborators, DUMAND proposed using the deep ocean as a vast Cherenkov detector for atmospheric and high-energy astrophysical neutrinos \citep{Roberts1992}. Although the project never achieved its original scientific objectives, it established many of the conceptual, technological, and organizational foundations upon which later underwater and under-ice neutrino telescopes were built \citep{Spiering2019}. In this sense, DUMAND represented a crucial step in transforming high-energy neutrino astronomy from a theoretical vision into a realistic experimental programme.\footnote{In 1980, the Reagan administration ended cooperation in response to the Soviet invasion of Afghanistan. At the same time, however, A. Chudakov proposed using the deep waters of Lake Baikal in Siberia for a  ``Russian DUMAND'' \citep{Spiering2019}.}

Parallel developments were also reshaping the relationship between particle physics and cosmology. The discovery of the cosmic microwave background radiation in 1965 demonstrated that relics from the early Universe could survive to the present epoch, providing the first direct observational evidence for physical processes occurring shortly after the Big Bang \citep{SciamaMandolesi1990}. Grand Unified Theories further transformed cosmology by predicting relic particles such as magnetic monopoles and stable particles predicted by emerging theories of fundamental interactions that might have survived from the earliest moments of the Universe. Like the cosmic microwave background, these relics were conceived as observable remnants of the hot Big Bang, but now carrying information about elementary-particle physics rather than only about cosmological evolution. While magnetic monopoles became important as potential signatures of Grand Unified Theories, stable relic particles increasingly attracted attention because they might account for part of the missing mass of the Universe. Within this framework, dark matter gradually evolved from a predominantly astronomical problem into a meeting ground between cosmology and elementary-particle physics \citep{BertoneHooper2018}.

The discovery of the cosmic microwave background also had important implications for studies of the highest-energy cosmic particles. In 1966, Kenneth Greisen and, independently, Georgii Zatsepin and Vadim Kuzmin showed that ultra-high-energy protons propagating over cosmological distances should interact with the background radiation, leading to substantial energy losses above a threshold energy \citep{Greisen1966} \citep{Zatsepin:1966jv}. Although the interpretation of the observed suppression of the ultra-high-energy cosmic-ray spectrum later proved more complex than originally envisaged, the GZK prediction represented a major conceptual advance. It demonstrated that relic radiation from the early Universe could directly influence high-energy particle propagation, providing one of the earliest examples of the growing interaction between cosmology, particle physics, and cosmic-ray physics. It also provided an upper limit for cosmic-ray energies. 

At the same time, the long-standing problem of cosmic-ray origins acquired renewed urgency. New astronomical discoveries revealed increasingly plausible sites of high-energy particle acceleration, yet charged cosmic rays could not readily identify their sources because magnetic fields erased directional information. This limitation stimulated growing interest in complementary observational channels capable of tracing high-energy particle accelerators directly. In particular, very-high-energy $\gamma$ rays offered the possibility of identifying candidate sources through directional observations, providing a major impetus for the development of atmospheric Cherenkov techniques.
These changing priorities were reflected in the rapid evolution of very-high-energy $\gamma$-ray astronomy. Atmospheric Cherenkov techniques, originally developed within cosmic-ray research, underwent major refinement through advances in instrumentation, image analysis, and increasingly sophisticated Monte Carlo simulations. In particular, the work of A. M. Hillas showed how image parameterization could discriminate $\gamma$-ray-induced air showers from the overwhelming hadronic background \citep{Hillas1984}, providing one of the key methodological advances that underpinned the modern development of very-high-energy $\gamma$-ray astronomy.

A particularly significant expression of these transformations was the proposal, put forward by Antonino Zichichi in 1979, for the Gran Sasso National Laboratory \citep{Zichichi1983}. Conceived from the outset as a multidisciplinary underground research centre, the project combined an unusually broad scientific programme, including proton-decay searches, neutrino physics, solar-neutrino studies, long-baseline accelerator neutrino experiments, cosmic-ray research, geophysics, and investigations of biologically active matter. Gran Sasso embodied a new conception of the underground laboratory: a permanent scientific infrastructure designed from the outset to support multiple complementary experimental programmes requiring exceptionally low-background environments.

By the end of the 1970s, neither a coherent astroparticle community nor a clearly defined astroparticle identity yet existed. Nevertheless, the scientific landscape had changed profoundly. Particle physicists increasingly looked beyond the accelerator, cosmic-ray physicists continued to broaden their observational approaches to high-energy phenomena, while astrophysical and cosmological questions became progressively intertwined with elementary-particle physics. The search for proton decay, neutrinos, $\gamma$ rays, dark matter, and the origin of cosmic rays increasingly created common scientific problems that brought previously distinct research traditions into sustained dialogue. In parallel, what Gary Steigman aptly described as cosmology ``confronting particle physics'' \citep{Steigman1979} increasingly became a defining feature of the emerging interaction between the two fields \citep{Barrow1982} \citep{Sciama1983}. These developments did not yet constitute a new discipline, but they established many of the conceptual, observational, experimental, and institutional conditions from which astroparticle physics would gradually emerge during the following decade. 

\section{The 1980s: Crystallization of a New Intellectual Landscape}
\label{Section 3}
By the early 1980s, many of the theoretical developments that had emerged during the previous decade began to coalesce into a new intellectual landscape linking particle physics, cosmology, and gravitation. Increasingly sophisticated theories of elementary particles were applied to the physics of the early Universe, consolidating the emerging field of particle cosmology. Grand Unified Theories further transformed cosmology by predicting relic particles that might have survived from the earliest moments of the Universe. Like the cosmic microwave background, these relic particles were understood as observable remnants of the hot Big Bang, but ones expected to preserve information about elementary-particle processes occurring in the earliest fractions of a second after the origin of the Universe. Dark matter thus progressively ceased to be merely an astronomical problem concerning the dynamics of galaxies and became one of the principal points of contact between cosmology and elementary-particle physics. Questions concerning baryogenesis, dark matter, neutrino properties, vacuum phase transitions, and the earliest fractions of a second after the Big Bang increasingly came to be addressed within a common theoretical framework. The early Universe itself increasingly came to be regarded as a natural laboratory for exploring physical conditions far beyond the reach of terrestrial accelerators. As Yakov Zel'dovich had already argued in 1970, the hot Big Bang could be viewed as a ``laboratory for the nuclear and particle physicist" and even as a ``poor man's accelerator", capable of probing energies inaccessible to laboratory experiments \citep{Zeldovich1970}.\footnote{For the historical origins of this conception within the broader emergence of relativistic astrophysics and the Soviet tradition, see \citep{FurlanBonolis2025}.} Rather than relying exclusively on laboratory experiments, particle physics increasingly turned to cosmological observations as a means of testing theories of fundamental interactions and constraining the properties of elementary particles. In this way, cosmology progressively evolved from a discipline primarily concerned with the large-scale structure and evolution of the Universe into a powerful source of empirical constraints on microphysics. Developments in black-hole physics, quantum fields in curved spacetime, and inflation further strengthened the interaction between particle physics, gravitation, and cosmology. One of the earliest and most influential manifestations of this new perspective was provided by Big Bang nucleosynthesis, whose successful predictions linked the observed primordial abundances of the light elements to the physical conditions prevailing during the first minutes after the Big Bang \citep{IoccoEtal2009}. By the late 1970s and early 1980s, comparisons between nucleosynthesis calculations and astronomical observations were already providing quantitative constraints on elementary-particle physics, most notably on the number of light neutrino species, demonstrating that cosmological observations could serve as powerful tests of particle-physics models \citep{Steigman1979}  \citep{Copi1997} \citep{SchrammTurner1998}. Big Bang nucleosynthesis thus transformed cosmology into a quantitative probe of elementary-particle physics \citep{Olive2000}. The cosmological limit on the number of light neutrino species inferred from Big Bang nucleosynthesis was independently confirmed by the first precision measurements of the invisible decay width of the $Z{^0}$ boson at LEP in 1989-1990, providing one of the earliest and most striking examples of the reciprocal interplay between cosmology and accelerator particle physics \citep{Steigman1979} \citep{ALEPH1989} \citep{L31990}. This remarkable agreement demonstrated that cosmological observations and accelerator experiments had become mutually constraining sources of knowledge about elementary particles, illustrating the new relationship between microphysics and cosmology that lay at the heart of particle cosmology.

These developments found a particularly influential expression in the 1982 Cambridge workshop The Very Early Universe, organized by Stephen Hawking and colleagues \citep{Hawking1983}. Bringing together particle physicists, cosmologists, and astrophysicists, the meeting helped consolidate an intellectual programme that had been taking shape throughout the previous decade around the interactions between microphysics and cosmology \citep{Steigman1984}. Questions concerning dark matter, baryogenesis, inflation, phase transitions, and the physics of the early Universe increasingly became part of a common research agenda. 
These theoretical developments were accompanied by an equally important process of intellectual consolidation. During the late 1970s and early 1980s, a number of physicists actively worked to transform these interconnected research directions into a coherent scientific programme.

Building on the pioneering vision of Yakov Zel'dovich, among the principal architects of this emerging research programme were David Schramm, Michael Turner, and Edward Kolb, whose influential reviews, conferences, edited volumes, and textbooks helped establish particle cosmology as a coherent area of research \citep{Zeldovich1970} \citep{TurnerSchramm1985} \citep{Kolb1988} \citep{SchrammTurner1998} \citep{KolbTurner1990}. By the mid-1980s, the programme envisaged by Zel'dovich had largely become a reality: the early Universe and high-energy astrophysical phenomena were increasingly recognized as natural laboratories for fundamental physics, while cosmological observations were beginning to provide empirical constraints on elementary-particle theories \citep{Zeldovich1988} \citep{TurnerSchramm1979} \citep{Kolb2007}.

A parallel process unfolded at the institutional level. The ESO--CERN Symposia represented a fundamentally different stage in this historical process \citep{Setti1984} \citep{Setti1987}. Whereas the Texas Symposia had shown that Nature itself could compel previously distinct scientific communities to confront common problems, the joint initiative of ESO and CERN reflected a conscious institutional recognition that this dialogue had become a permanent feature of frontier research. The joint sponsorship by the European Southern Observatory and CERN---the two European institutions most closely associated with observational astronomy and high-energy particle physics, respectively---was historically significant in itself. The very fact that the meetings were jointly organized by these institutions symbolized the growing realization that many of the fundamental questions concerning the Universe could no longer be addressed within the traditional boundaries of either discipline alone. Rather than merely providing a forum for scientific exchange, the symposia constituted an explicit institutional effort to foster sustained collaboration between communities whose interactions had previously been occasional rather than structural. In this respect, the ESO--CERN Symposia did not simply reflect an emerging interdisciplinary landscape; they actively contributed to shaping it.

At the same time, the long-standing problem of the origin of cosmic rays continued to stimulate new observational strategies. $\gamma$-ray astronomy, neutrino astronomy, and extensive air-shower experiments increasingly formed a complementary observational framework for investigating the acceleration and propagation of high-energy particles and the nature of their astrophysical sources. Together, these developments broadened the observational repertoire available for investigating the high-energy Universe. More broadly, they revealed an increasingly coherent scientific landscape in which new theoretical frameworks, observational techniques, and institutional initiatives encouraged sustained interaction among communities that had previously evolved along separate trajectories. By the mid-1980s, many of the conceptual, methodological, and institutional foundations of the emerging field were firmly in place.

\section{From Particles to Messengers}
\label{Section 4}
By the late 1980s, the scientific developments described in the preceding sections had reached a new level of coherence. Research programmes that had evolved along largely independent trajectories---including underground rare-event physics, neutrino and $\gamma$-ray astronomy, particle cosmology, and cosmic-ray physics---increasingly addressed common scientific problems through complementary experimental approaches. Astroparticle physics became recognizable not simply through a series of major discoveries, but through the emergence of a shared scientific agenda, increasingly interconnected experimental practices, and new observational strategies capable of linking elementary-particle physics with the high-energy Universe. 

The convergence of the theoretical, experimental, and institutional developments outlined in the preceding sections found its first spectacular experimental validation in the detection of neutrinos from Supernova 1987A. On 23 February 1987, bursts of neutrinos associated with the explosion of a supernova in the Large Magellanic Cloud were recorded by underground detectors in Japan, the United States, and the Soviet Union \citep{Hirata1987} \citep{Bionta1987} \citep{Alekseev1987}. For the first time, neutrinos originating beyond the Solar System had been directly observed. The detection confirmed key aspects of stellar-collapse theory while simultaneously demonstrating that astrophysical neutrinos could serve as probes not only of otherwise inaccessible stellar interiors but also of fundamental neutrino properties, including their mass, interactions, and possible physics beyond the Standard Model. More broadly, Supernova 1987A represented the first spectacular validation of an experimental strategy that had been taking shape over the preceding decades: the use of naturally occurring particles not only as astrophysical messengers but also as probes of fundamental physical processes \citep{Bonolis2025} \citep{LaRana2025}.

The historical significance of SN1987A extended well beyond its immediate scientific results. The explosion prompted an unprecedented international observational  campaign, spanning optical, radio, ultraviolet, X-ray, and neutrino astronomy. Although it did not yet represent multi-messenger astronomy in its modern coordinated form, it clearly demonstrated the scientific value of combining complementary observational channels to investigate a single astrophysical phenomenon. In this respect, SN1987A represented the culmination of a long process during which the multiwavelength exploration of the high-energy Universe progressively merged with an emerging messenger-based perspective. More generally, neutrinos had progressively become one of the principal connecting elements of the emerging field. They linked underground cosmic-ray experiments with stellar physics, supernova explosions, and, through Big Bang nucleosynthesis and dark-matter scenarios, with particle cosmology itself.
By combining observations across the electromagnetic spectrum with the first detection of neutrinos from beyond the Solar System, SN1987A marked a decisive milestone in the transition from the multiwavelength exploration of the high-energy Universe to the coordinated use of different astronomical messengers. Although multimessenger astronomy became an established observational paradigm only during the following decades, particularly through coordinated campaigns on transient phenomena such as $\gamma$-ray bursts and later with the addition of gravitational-wave observations, SN1987A provided its first compelling demonstration \citep{Lalli2025}.

The late 1980s also witnessed the realization of scientific infrastructures that gave concrete expression to the emerging astroparticle landscape. Among the most significant was the Gran Sasso National Laboratory in Italy, brought into operation during the presidency of Nicola Cabibbo at INFN, with Enrico Bellotti serving as its first Director. The Laboratory brought together research programmes that had previously developed within largely separate scientific traditions. Experiments such as EAS-TOP, GALLEX, LVD, and MACRO demonstrated how previously distinct experimental traditions and scientific objectives could coexist and develop within a shared research environment \citep{Zanotti1991}.
Equally important was the scientific culture that developed within the laboratory. From its inception, Gran Sasso was deliberately organized as an international research institution, bringing together Italian groups and major collaborations from the Soviet Union, Germany, the United States, France, Israel, and other countries. These communities arrived with different experimental traditions, institutional practices, and scientific priorities, yet increasingly interacted through shared infrastructures, common technical challenges, and complementary experimental programmes. In this respect, Gran Sasso became not only a place where diverse experiments coexisted, but also an environment in which shared experimental challenges and common scientific objectives progressively fostered the interaction of previously distinct scientific cultures.

This strategy proved remarkably successful. The first underground detections of atmospheric neutrinos in India and South Africa during the 1960s had established a new class of naturally occurring particles available for experimental study. Building upon this foundation, increasingly sophisticated underground experiments during the following decades revealed anomalies in the atmospheric-neutrino flux that ultimately led to the discovery of neutrino oscillations. At the same time, solar-neutrino observations provided direct evidence that the Sun is powered by thermonuclear reactions while simultaneously revealing the long-standing solar-neutrino deficit. The eventual interpretation of both atmospheric and solar neutrino anomalies in terms of neutrino oscillations demonstrated that neutrinos possess non-zero masses, providing one of the first clear indications of physics beyond the Standard Model. Naturally occurring particles were no longer merely objects of investigation; they had become indispensable tools for exploring both astrophysical processes and the fundamental laws of nature.

An equally significant transformation occurred in very-high-energy $\gamma$-ray astronomy. In Europe, this evolution became particularly visible through the HEGRA collaboration. Emerging from the Kiel cosmic-ray tradition and stimulated in part by the scientific debate surrounding the Cygnus X-3 episode---which, despite its ultimately incorrect interpretation, stimulated important methodological advances and attracted many particle physicists to very-high-energy $\gamma$-ray astronomy---HEGRA introduced stereoscopic atmospheric Cherenkov observations, allowing the same air shower to be imaged simultaneously by several telescopes. This represented a decisive methodological advance, substantially improving event reconstruction, angular resolution, energy determination, and discrimination between $\gamma$-ray- and hadron-induced showers.

The excitement generated by the Cygnus X-3 observations catalyzed the entry of a new generation of experimental particle physicists into very-high-energy $\gamma$-ray astronomy. Among them were Eckart Lorenz and Werner Hofmann, whose experience in CERN detector experiments introduced technologies, data-analysis methods, and collaborative practices developed in accelerator physics. These competencies converged within the large international HEGRA collaboration, where they interacted with the long-standing cosmic-ray tradition represented by Otto Claus Alkofer's Kiel group, the theoretical expertise in non-thermal astrophysics and plasma physics associated with Heinrich V{\"o}lk and the Max Planck Institute for Nuclear Physics in Heidelberg, and the important contributions of Razmik Mirzoyan, Felix Aharonian, and many other scientists from the former Soviet Union and across Europe. HEGRA thus represented not only a successful experiment, but also a transitional stage in the development of European very-high-energy $\gamma$-ray astronomy. Emerging from the long tradition of European cosmic-ray research, it illustrates particularly clearly how experimental techniques, detector technologies, and scientific questions were progressively transferred from cosmic-ray physics to $\gamma$-ray astronomy, laying the foundations for later observatories such as MAGIC and H.E.S.S. \citep{BonolisLeon2023}. 

In parallel, the Whipple tradition in the United States evolved into VERITAS, illustrating how complementary experimental lineages on both sides of the Atlantic converged toward the modern generation of imaging atmospheric Cherenkov observatories \citep{Hillas2013} \citep{Fegan2019}. Taken together, these developments showed that the transformations initiated during the previous decades had reached maturity. Underground laboratories and atmospheric Cherenkov observatories had demonstrated their scientific potential, while previously distinct experimental traditions increasingly shared common scientific questions, observational strategies, and experimental practices. Although the institutional consolidation of astroparticle physics would largely occur during the following decade, its principal scientific foundations had by then been firmly established.

\section{Forging an International Community}
\label{Section 5}
By the early 1990s, the scientific transformations described in the previous sections increasingly found institutional recognition. Research situated at the intersection of particle physics, astrophysics, and cosmology was no longer regarded simply as a collection of related activities but as an emerging field requiring dedicated scientific forums, funding structures, and international coordination. This institutional recognition unfolded simultaneously at national and international levels. 

In the United States, the 1991 Decadal Survey \citep{NRC1991} established the first dedicated panel on Particle Astrophysics, formally acknowledging the growing importance of research programmes situated between particle physics, astrophysics, and cosmology \citep{Bahcall1991} \citep{ParticleAstrophysicsPanel1991} \citep{Bahcall1991WorkingPapers}. 

Similar developments occurred in Europe, where underground laboratories, cosmic-ray observatories, and emerging $\gamma$-ray and neutrino projects increasingly attracted coordinated support from national research agencies and European scientific organizations. Alongside CERN and ESO, the European Science Foundation progressively provided a framework for transnational scientific coordination, reflecting a growing awareness that many of the most ambitious programmes at the intersection of particle physics, astrophysics, and cosmology required shared infrastructures, long-term planning, and sustained international collaboration. These developments anticipated the more formal organizational structures that would emerge during the 1990s and early 2000s. During the same period, dedicated research centres, university groups, and graduate programmes devoted to astroparticle physics also began to appear, providing further evidence that the field was acquiring a stable institutional identity.

The historical transformations described in the preceding sections found a parallel expression in the progressive emergence of new institutional structures. Conferences, journals, coordinating bodies, and strategic research networks did not merely accompany this transformation; they progressively gave institutional form to an already evolving scientific landscape.
 
The changing character of the field was equally reflected in scientific meetings and educational initiatives. Long-established forums such as the International Cosmic Ray Conferences and the Texas Symposia continued to foster exchanges among communities that had historically evolved along different trajectories. A further step occurred with the First International School on Astro-Particle Physics, held in Erice in 1987 under the auspices of CERN and ESO \citep{DeRujula1987}. The School's programme explicitly declared that ``recent progress in particle physics, cosmology and astrophysics has given birth to a discipline that encompasses them all" and stressed the need for an interdisciplinary treatment of these developments. Significantly, the School built upon the dialogue initiated by the two ESO--CERN Symposia of 1983 and 1986, thereby marking not so much the beginning of astroparticle physics as the culmination of a long period of emergence during which previously distinct research traditions had progressively become interconnected. The Erice School was certainly the first of this kind under the new label astro-particle physics, showing that the field was beginning to be recognized as a common research domain for communities that had previously evolved along separate disciplinary trajectories.\footnote{To the best of my knowledge, the earliest printed occurrence of the term astro-particle is due to Gary Steigman, who used it in a review article published in Nature in 1984 \cite{Steigman1984}. The expression astro-particle physics appeared shortly afterwards in the title of a paper by Abdus Salam presented at the Fourth Marcel Grossmann Meeting in 1985 \citep{Salam1986}. Although the terminology had thus entered the literature somewhat earlier, the Erice School appears to have been the first international scientific meeting explicitly organized under the label Astro-Particle Physics, signalling the emergence of the term as the designation of a new interdisciplinary research field.}

Two years later, the launch of the conference series Topics in Astroparticle and Underground Physics (TAUP) in 1989 explicitly acknowledged the convergence of underground physics, neutrino research, dark-matter searches, cosmic-ray studies, and high-energy astrophysics. Dedicated schools, topical workshops, and interdisciplinary conferences further strengthened exchanges among researchers from particle physics, astrophysics, cosmology, and cosmic-ray physics. This transitional phase was also reflected in the contemporary literature. Volumes such as \textit{Particle Astrophysics and Cosmology}, edited by Maurice Shapiro, Rein Silberberg, and John Wefel in 1993, brought together contributions on cosmic rays, $\gamma$ rays, neutrinos, and cosmology, illustrating the growing recognition that these subjects were becoming components of a common scientific enterprise even before the institutional consolidation of astroparticle physics. Reflecting this changing intellectual landscape, Maurice Shapiro observed that ``the symbiosis between particle physics and cosmology'' had virtually become ``a conjugal relationship'' \citep[p.~vii]{Shapiro1993}. The metaphor captured more than the growing interaction between two disciplines; it captured the wider reconfiguration of high-energy research in which astronomy, cosmology, particle physics, and astrophysics were increasingly drawn together by common scientific problems.

The emergence of dedicated publication venues and international coordinating structures further consolidated this process. In Europe, these developments were accompanied by a progressive strengthening of transnational scientific coordination. Building upon the long-standing collaborative philosophy of the European Science Foundation, increasingly ambitious astroparticle projects during the 1990s required new forms of strategic planning and multinational cooperation, complementing the collaborative roles already played by organizations such as CERN and ESO. This institutional evolution reflected---and helped consolidate---the scientific emergence of astroparticle physics reconstructed in the present article.\footnote{The development of European coordination in astroparticle physics formed part of a broader evolution in European science policy. Since its foundation in 1974, the European Science Foundation (ESF) sought to promote transnational scientific cooperation and to complement national research organizations through collaborative programmes and strategic initiatives rather than direct funding, an approach clearly articulated by its first President, Brian Flowers \citep{Flowers1978}. During the 1990s, following the end of the Cold War, the ESF increasingly emphasized scientific networking, coordinated research programmes, and long-term strategic planning for large-scale research infrastructures, as reflected in the ESF 1997 Annual Report and the preparation of its 1998–2001 Action Plan \citep{ESF1997CORDIS}. In parallel, European support for astroparticle physics was already taking shape through dedicated collaborative initiatives such as the Theoretical Astroparticle Network (TAN) (1994–1997), established within the European Union's Human Capital and Mobility Programme, which brought together leading European institutions including CERN, INFN, CNRS, the Max Planck Society, Oxford University, Nordita and others \citep{TAN1994}. These initiatives anticipated the creation of the ESF Astroparticle Physics Network in the late 1990s and the establishment of the Astroparticle Physics European Coordination (ApPEC) in 2001. The subsequent ASPERA and ASPERA-2 ERA-NET programmes (2006–2012) further strengthened European coordination by developing common roadmaps, joint funding mechanisms, and strategic planning, paving the way for the establishment of the present Astroparticle Physics European Consortium (APPEC) in 2012. } 

These initiatives reflected the growing recognition that many of the scientific challenges at the intersection of particle physics, astrophysics, and cosmology could only be addressed through coordinated international efforts. The launch of the journal \textit{Astroparticle Physics} in 1992, followed a decade later by the \textit{Journal of Cosmology and Astroparticle Physics} (JCAP), provided common publication forums for research previously dispersed across particle-physics, astrophysical, cosmological, and cosmic-ray journals. 

At the same time, the growing scale and complexity of underground laboratories, large cosmic-ray arrays, $\gamma$-ray observatories, and proposals for high-energy neutrino telescopes made international coordination increasingly indispensable. Within the framework of the International Union of Pure and Applied Physics (IUPAP), the traditional structures historically centred on cosmic-ray research progressively expanded to accommodate the broader scientific landscape. This evolution culminated in the establishment of the Particle and Nuclear Astrophysics and Gravitation International Committee (PaNAGIC) in 1998, which broadened the international coordinating role that had long been associated with Commission 4 on Cosmic Rays. By bringing together particle physics, astrophysics, cosmology, gravitation, and related fields within a common organizational framework, PaNAGIC formally acknowledged the increasingly interdisciplinary character of the field. Its dedicated working groups on high-energy neutrinos and gravitational waves also anticipated research directions that would later become central to multi-messenger astronomy.\footnote{See Report of PaNAGIC to IUPAP, presented at the meeting of the Council and Commission Chairs of IUPAP, Beijing, 6 October 2000, Revised November 2000: \url{https://archive.iupap.org/wg/wg4/}.}

These efforts culminated in the establishment of the Astroparticle Physics European Coordination (ApPEC) in 2001, which provided a common framework through which European funding agencies could coordinate scientific priorities and long-term planning for the emerging field \citep{Spiering2008}.\footnote{The present Astroparticle Physics European Consortium (APPEC) was established in 2012. It developed from the Astroparticle Physics European Coordination committee (ApPEC), created in 2001, following the preparatory work carried out within the ASPERA and ASPERA-2 ERA-NET programmes (2006-2012) (see the Final Report Summary: \url{https://cordis.europa.eu/project/id/235489/reporting/es}. Among their principal achievements was the publication of the first \textit{European Roadmap for Astroparticle Physics} (2008) (\url{https://www.uv.es/aspera/RoadmapPhase1.pdf}) which was prepared between October 2005 and January 2007. As specified on p. 3, the need for such a roadmap arose since projects in astroparticle physics were moving to ``ever larger sensitivity and scale, with costs of individual projects on the 100 MEuro scale or beyond''. The document identified the major scientific priorities and large-scale infrastructures required for the future development of the field while establishing common mechanisms for strategic planning and transnational cooperation. Significantly, its thematic organization---covering cosmology and the early Universe, particle properties, neutrinos, the non-thermal Universe, and gravitational waves---reflected the historical transformations reconstructed in the present article.} Together with IUPAP and PaNAGIC, ApPEC reflected the growing institutional maturity of astroparticle physics as an internationally organized research field \citep{Bettini2025}. 

By the turn of the twenty-first century, astroparticle physics possessed many of the characteristics associated with a mature international research field. Dedicated journals, specialized conferences, representative organizations, coordinated planning exercises, and increasingly ambitious experimental infrastructures provided stable frameworks within which researchers pursued common scientific objectives. 

Its first successful achievements were the opening of the neutrino window through the detection of neutrinos from the Sun---which provided the first clear evidence that neutrinos have mass---and from Supernova 1987A. The first unambiguous detection of TeV $\gamma$ rays from the Crab Nebula by the Whipple collaboration in 1989 validated the imaging atmospheric Cherenkov technique and marked the beginning of modern ground-based very-high-energy $\gamma$-ray astronomy \citep{Weekes1989}. The 1989 detection of the Crab Nebula demonstrated the viability of the imaging atmospheric Cherenkov technique, while subsequent European experiments---most notably HEGRA---transformed this proof of principle into a mature observational methodology through stereoscopic imaging, ultimately laying the foundations for H.E.S.S. and MAGIC \citep{Mirzoyan2022}. By the mid-2000s, ground-based Cherenkov observatories had detected several tens of very-high-energy $\gamma$-ray sources, demonstrating that $\gamma$-ray astronomy had evolved into a fully established observational discipline and providing one of the principal experimental pillars of the emerging astroparticle paradigm.

In 2008, the first European Roadmap for Astroparticle Physics  identified six major fundamental questions for the field's development over the next decade: 

1) What is the Universe made of? In particular: What is dark matter?

2) Do protons have a finite life time?

3) What are the properties of neutrinos? What is their role in cosmic evolution?

4) What do neutrinos tell us about the interior of the Sun and the Earth, and
about Supernova explosions?

5) What is the origin of cosmic rays ? What is the view of the sky at extreme
energies ?

6) Can we detect gravitational waves ? What will they tell us about violent cosmic
processes and about the nature of gravity? 

Towards the end of the decade, the observational and experimental capabilities required to address many of these questions had reached an unprecedented level of maturity. The spectacular detections of astrophysical neutrinos and gravitational waves therefore represented the culmination of historical developments that had been unfolding since the postwar transformation of cosmic-ray physics. In this sense, the detection of the neutron-star merger GW170817 \citep{AbbottEtAl2017} did not inaugurate an entirely new historical trajectory, but rather marked the culmination of a much longer process through which astroparticle physics, itself the product of the historical transformations reconstructed in this article, progressively became integrated with multiwavelength and, subsequently, gravitational-wave astronomy into the multimessenger observational paradigm.

\section*{Conclusion}
The historical reconstruction presented in this article invites a broader reconsideration of how astroparticle physics emerged as a distinct scientific field. Its emergence was not the outcome of a sudden scientific revolution or the simple convergence of previously established disciplines, but of a long historical process during which experimental traditions, scientific questions, and research communities that had evolved largely independently progressively became interconnected. By following this process from the transformation of postwar cosmic-ray physics to the institutional consolidation of the field during the 1990s, this article has argued that astroparticle physics emerged through the gradual reconfiguration of experimental cultures rather than through the replacement of existing disciplinary boundaries.

Three broad historical trajectories shaped this development. The transformation of cosmic-ray physics redirected attention from the discovery of elementary particles toward the origin, acceleration, and propagation of high-energy cosmic particles, while preserving an experimental culture based on naturally occurring messengers, geographically distributed observatories, and the exploitation of natural environments as components of the experimental apparatus. At the same time, the emergence of relativistic astrophysics and the new astronomical windows opened by radio, X-ray, $\gamma$-ray, and neutrino observations revealed a Universe populated by compact objects, relativistic plasmas, and powerful natural accelerators. Finally, the growing interaction between particle physics and cosmology established new connections between microphysics and the early Universe, transforming cosmological observations into probes of fundamental physics. Rather than following separate trajectories, these developments progressively interacted through common scientific questions, complementary observational strategies, and, in some cases, shared observational messengers. Among these, neutrinos occupied a unique position by linking cosmic-ray physics, solar, stellar and high-energy astrophysics, and particle cosmology within a common experimental and theoretical framework.

The analysis developed in this article suggests that the emergence of astroparticle physics was driven less by the convergence of established disciplines than by the appearance of scientific problems that could no longer be addressed within their traditional boundaries. The search for the origin of cosmic rays, the detection of rare events, the investigation of dark matter, the study of neutrinos, and the exploration of the early Universe all demanded new experimental practices, novel infrastructures, and increasingly international forms of collaboration. In this sense, scientific problems acted as powerful agents of reorganization, reshaping experimental cultures and bringing previously distinct communities into sustained interaction.

This perspective also helps explain why the institutional recognition of astroparticle physics occurred relatively late. Dedicated journals, conferences, laboratories, research centres, university groups, and international organizations progressively gave institutional form to a scientific landscape that had already taken shape through decades of intellectual exchange, methodological innovation, and experimental collaboration. Institutional consolidation therefore represented the culmination of a much longer process of scientific and organizational transformation.

More broadly, the historical reconstruction proposed here illustrates a characteristic feature of the development of modern science. New scientific fields do not necessarily arise through the simple merger of existing disciplines, but through the progressive reconfiguration of research traditions around shared scientific problems. As these problems increasingly transcend established intellectual and institutional boundaries, they reshape scientific questions, experimental practices, research communities, and research infrastructures, thereby giving rise to new scientific fields. The subsequent development of dedicated publication venues, scientific forums, coordinating bodies, and funding frameworks does not initiate this process but recognizes, stabilizes, and further enables the emerging field.

As new research frontiers emerged beyond the reach of conventional experimental methods, they progressively reshaped not only experimental strategies and scientific institutions but also the careers of researchers themselves. Physicists trained in accelerator-based particle physics increasingly entered astroparticle research, transferring detector technologies, organizational models, and experimental cultures to the investigation of naturally occurring high-energy phenomena in relationship with particle physics problems such as neutrino mass, proton decay, search for WIMPS and other exotic entities or nucleosynthesis processes in different astrophysical environments. Rather than replacing accelerator-based research, these new frontiers complemented and extended it, demonstrating how Nature itself could become a laboratory for fundamental physics. The history reconstructed here therefore illuminates not only the origins of astroparticle physics, but also the broader mechanisms through which contemporary science continually reorganizes scientific communities and creates new ways of investigating nature.

\section*{Acknowledgments}
I would like to express my gratitude to Alessandro Bettini, Alessandro De Angelis, Enzo Iarocci and Francesco Vissani for their insightful comments on a preliminary draft of the manuscript.

\bibliographystyle{apalike-refs}
\bibliography{Bibliography_Astro_Proceedings}

\end{document}